\documentclass[11pt]{article}

\usepackage[utf8]{inputenc} 
\usepackage{amsmath, amssymb} 
\usepackage{amsthm}
\usepackage{graphicx} 
\usepackage{hyperref} 
\usepackage{authblk}  
\usepackage{geometry} 
\usepackage{subcaption}
\usepackage{quotes}
\title{ Consistent control of drying rates of solution thin films  on wafer-sized substrates by dynamic air-knife drying with optimal trajectories}

\author[1]{Simon Ternes}
\affil[1]{Department of Electrical Engineering, University of Rome “Tor Vergata”, via del Politecnico 1, Rome, 00133 Italy}
\newtheorem*{theorem*}{Optimization Goal}
\date{\today}

\begin{document}

\maketitle

\begin{abstract}

This work tackles the problem of achieving consistent drying rates of a solution film deposited on a $20\,\rm{cm}$-wide substrate ($\approx $ silicon-wafer size) that is driven under a narrow air flow ejected by a slot nozzle (or "air knife"). The main prerequisite of the work is that the drying rate of the solution film is highly decisive for a certain performance indicator of the deposited film at a particular, critical concentration $c_{\rm crit.}$. Empirically, this concentration can  be  associated with the visual observation of "the drying front" as, for the example of hybrid perovskite thin films, caused by the onset of a crystallization process. As a main result, a set of equations for achieving consistent drying rates, $\dot{d}_{\rm crit.}$, at critical concentration is presented that is solved by a simple two-staged least-squares gradient decent.  From the resulting velocity vector, an optimal trajectory of the air knife, $\hat{x}(t)$, depending on the initial wet film thickness distribution over the substrate is derived. It is demonstrated that scenarios where the wet film thickness increases along the movement direction of the air knife have a consistent set of equations. Wet thin films that do not obey this constraint, as in the demonstrated scenarios with convex and concave shapes of wet film thickness over the substrate area, cannot always be dried in a fully consistent way by optimizing the air-knife trajectory alone. However, with the presented methods, optimal trajectories can still be derived that enable more homogeneous drying results.
\end{abstract}

\section{Introduction}
The industrial process of coating and drying thin films is an essential part of many applications ranging from photographic film to varnishes and paintings\cite{Heyd25,buttThinFilmCoatingMethods2022,niskanenMonitoringDryingProcess2019}. Most recently, thin film coating and drying is increasingly employed in technologies of the energy sector and electronics, such as batteries\cite{clementRecentAdvancesPrinted2022a}, fuel cells\cite{choolaeiRecentAdvancesChallenges2023}, light emitting diodes\cite{verbovenPrintingFlexibleLight2021}, thermoelectrics\cite{toshimaRecentProgressOrganic2017} and solar cells\cite{zhangRecentProgressEmerging2022,howardCoatedPrintedPerovskites2019}. There are two different industrial production modes for thin film coatings, which predetermine the way drying needs to be performed: Batch and continuous operation. In batch operation, individual substrates or batches of substrates are coated and dried successively\cite{fanEfficientPreciseSolutionvacuum2025}. This operation type is mostly employed when using rigid substrate materials and is also known as "Sheet-to-Sheet". In continuous processing of thin films, a long roll of a flexible substrate is continuously transferred from one roll to another ("Roll-to-Roll"), transitioning through one or multiple coating and/or drying steps\cite{sondergaardRolltoRollFabricationLarge2013}. The Roll-to-Roll configuration has the advantage that it is both economical and highly scalable\cite{parvazianRolltorollRevolutionTackle2024}. In addition, both the drying and the coating dynamics stabilize over the substrate area in case certain constraints on coating and drying parameters are fulfilled\cite{dingReviewOperatingLimits2016}. However, if rigid or fragile substrate units such as silicon wafers are processed, batch coating and drying might be the only option. When using high-throughput solution deposition such as slot-die coating or inkjet printing in batch operation, edge effects causing irregularities in the solution film thickness distribution can have an important impact on  product quality and yield \cite{parkPracticalOperationsIntermittent2022,diehmEdgeFormationHighSpeed2020,schackmarPerovskiteSolarCells2021}. These artifacts can further be exacerbated by the difficulty to dry these inhomogeneously  deposited wet films consistently on a limited substrate area\cite{geistertControllingThinFilm2023a}. In fact, even under the hypothesis of perfectly homogeneous coatings, consistent drying can pose a challenge. The reason is that gas nozzles or “air knives” employed for drying typically introduce spatial inhomogeneities in drying speeds over the substrate area\cite{ternesDryingCoatingPerovskite2022}. These homogeneity issues become  more pronounced with increasing air flow velocities. So, a mitigation strategy could be to favor  drying at lower air flow speeds. However, this is not always an option because of 1) a drying-rate dependent morphology formation requiring higher air flows (as was shown for example for perovskite PVs\cite{ternesDryingDynamicsSolutionProcessed2019,majewskiSimulationPerovskiteThin2025}) and/or 2) the necessity to increase production speeds for high throughput fabrication. Another possible mitigation strategy is to dynamically move the substrate with respect to the airflow during drying\cite{ternesDryingCoatingPerovskite2022}. Periodic movements, for example, effectively average the drying speeds over the substrate area such that different parts of the substrate dry with approximately the same speeds. However, if a drying-rate dependence of the morphology formation exists, the drying rate at different moments in time might be more decisive than at other moments in time, such that, despite the movement, the morphology formation will still depend on the drying rate present at these particular times\cite{majewskiSimulationPerovskiteThin2025,ternesDryingCoatingPerovskite2022,stargerFormationTrajectoriesSolutionprocessed2025}. For example, when depositing perovskite thin films, the drying rate (or supersaturation rate) at the time of crystallization onset is critical for forming morphology\cite{ternesModelingFundamentalDynamics2024a}. For these or similar material systems, drying should always be performed in a way that the most critical part of morphology formation is induced at a similar drying rate.

This work addresses the problem of  drying a previously deposited wet thin film on $20\,\rm{cm}$ long substrate ($\approx$ the size of an 8 inch silicon wafer) with a linearly moving air-knife exerting inhomogeneous mass transfer. The author has predominantly the application of organic and hybrid semi-conductor thin films in mind, but all the results can also be applied to other fields mentioned above when adjusting film thicknesses and the solvent parameters accordingly. The only prerequisite is that there is a certain critical  concentration of the dried solution (or dispersion) at which the morphology formation is primarily controlled by means of the drying rate. The movement trajectory of the air knife is then optimized to homogenize the  drying rate over the lateral substrate width at this particular stage of drying. The presented results translate directly to an experimental setup where the air knife or the substrate are positioned on a robot axis that allows for the setting of time-dependent acceleration (or jerk) values to follow the required trajectory over time. For modeling the drying dynamics on different positions of the substrate, an empirically test equation is borrowed from previous work on solution processing of perovskite solar cells \cite{ternesDryingDynamicsSolutionProcessed2019}. In detail, the investigated material system in this work was methylammonium lead iodide perovskite dissolved in N,N-Dimethylformamide (DMF). However, as alluded to above, the presented work is general in a sense that the parameters can be easily adjusted to any thin film drying process. The deposited thin film thickness distribution over the substrate area is, at first, assumed to be constant, and then, varied over the substrate area as it would be expected in a real-world coating setup. The work starts by optimizing the air knife trajectory in four scenarios of non-decreasing thin film thickness and demonstrates that these are, in general, consistently solvable. In eight successive scenarios of concave and convex coatings, it is showcased which start conditions lead to inconsistent drying in parts of the substrate area. Still, also for these scenarios, the dynamic air knife drying at optimized trajectory that approximately solves the proposed equations yields an immense improvement over classical drying at constant air-knife speed. In the future, the here-presented findings can be used to boost throughput and improve control, homogeneity and quality of batch drying processes by providing detailed insights into the entanglement of drying dynamics and substrate/air knife movement.

\section{Problem Statement and Methodology }

\subsection{Modeling of the drying thin film}

 In chemical engineering, the process of drying films was treated extensively on a theoretical basis by describing heat and mass transfer\cite{guerrierDryingKineticsPolymer1998,Heyd25} Typically, the drying process is divided into three sub-systems that are connected by means of their boundary conditions\cite{Ternes2023}: i) The diffusion of one or several liquid species within the film to the film surface, ii) the thermodynamic phase transition from the liquid to the gaseous phases of these species at the film surface and iii) the diffusion of the gaseous species into the surrounding gas volume. This report focuses on drying of thin film systems where i) is sufficiently fast throughout the drying process and ii) is always at thermodynamic equilibrium between condensation and evaporation, such that iii) limits the speed of solvent evaporation\footnote{The comparison of diffusion coefficients of i) and iii) restricts the validity of the equations used herein to either thin (not more than several $\mu$m) or dilute films (in more precise terms, the  analogous mass transfer Biot number must be smaller than 1). It is however possible  to include film-related drying effects in the optimization in the future.}.  The drying model applied herein is based on a starkly simplified linear equation of mass flux from \cite{ternesDryingDynamicsSolutionProcessed2019} that was empirically tested in drying perovskite solution thin films (details follow in the next section). For estimating the mass transfer in part iii), the spatially resolved Sheerwood-Number is used as detailed in section \ref{sec:mass_transfer}.

\begin{figure}[h]
\centering

\includegraphics[width=0.6\textwidth]{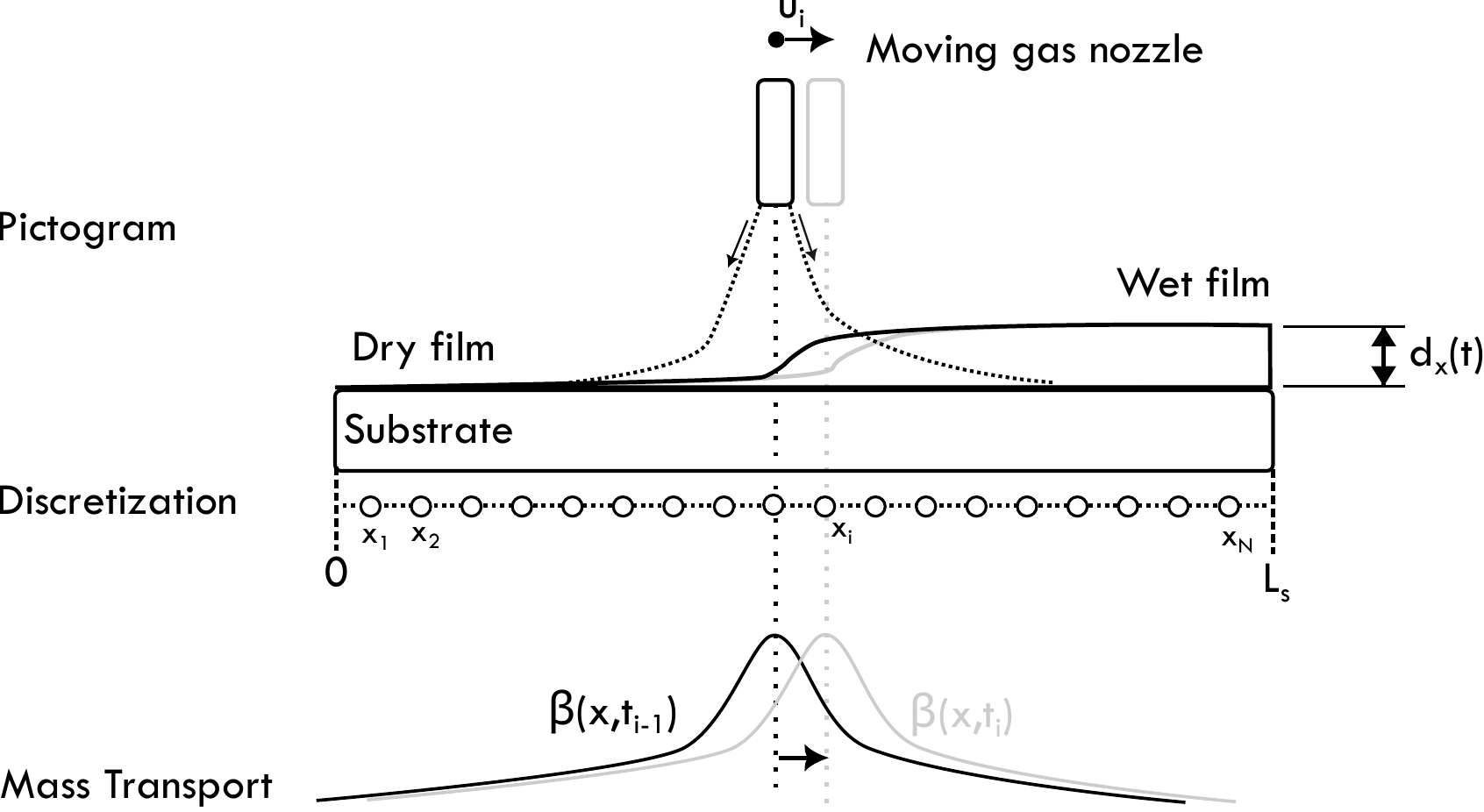}

\caption{Schematic depicting the one-dimensional, idealized problem addressed in this work and illustration of the one-dimensional spatial discretization employed. An air knife is moved with a certain speed $u_i$ from position $x_{i-1}$ to $x_i$ while exerting the gaseous mass transfer coefficient $\beta_{i}(t)$ on the film at this position. Evidently, the film will be dried fastest directly under the air knife, while the drying rate falls of toward the edges of the substrate.}
\label{fig:sample}
\end{figure}
 
\subsection{Dynamic air-knife drying on wafer-sized substrates}

 The idealized, mathematical problem addressed in this work is depicted in Figure \ref{fig:sample}. It was constructed to resemble batch drying as used in thin film electronics. For numerical discretization, the film is dissected into $N$ segments that are treated independently of each other\footnote{This is, strictly speaking, an assumption. It only holds if 1) lateral flow of solution and deformation of the film surface as possibly caused the air pressure can be neglected and 2) the mass transfer from the film is always active i.e. the concentration boundary layer is at all times aligned with the hydrodynamic boundary layer. 1) is plausible for films that are very thin, increase in viscosity when dried and exhibit strong adhesion to the substrate surface. 2) is justifiable because it is assumed that, in the ideal case, the film reaches its dried state approximately under the nozzle center.}. The total substrate length is set to $L_S=20\,\rm{cm}$, which is roughly the size of an 8 inch silicon wafer. Note that $L_S$ is an adjustable variable and only the ratio of the mass transfer coefficient's spatial dependence, $\beta(x)$, and the substrate size is decisive (details on $\beta(x)$ follow in the next section). For very large $L_S$ where $\beta(L_s) \ll \beta_{\max}$, the drying process will stabilize naturally along the axis of movement of the air nozzle and edge effects of inconsistent drying are less impactful (this  resembles approximately a Roll-to-Roll coating setup). However, if $L_S\ll \rm{FWHM} (\beta)$, the induced drying rate is naturally homogeneous and there is no need for further optimization as described herein. Therefore, this work applies to a situation where $L_m$ is on the order of tens of $\rm{FWHM}(\beta)$ and $\beta(L_s)$ is not neglectable in magnitude.  The air knife with the mass transfer $\beta(x)$  is then driven from the left to the right on this modeled film following the trajectory $\hat{x}(t)$. Note that, in experimental setups, the air knife is often static and the substrate is moved, but for purpose of clearer visualization, herein, $u_i$ is designated as  the “air knife velocity” from position $x_{i-1}$ to position $x_i$. However, the results can be applied to moving substrates in the same way as long as the same relative trajectory is present.  As a consequence of the dynamically moving air knife, every position, $x_i$, on the sample will experience a time-dependent $\beta _i(t)=\beta(x_i-\hat{x}(t))$ that is a function of the movement trajectory $\hat{x}(t)$. This mass transfer will, in turn, decrease the thickness of the thin film $d_{i}(t)$ at this position. The main idea of this work is to investigate how to control the trajectory $\hat{x}(t)$ of the air knife during the drying of the thin film such that a consistent result over all positions is reached. Consistency is measured  by evaluating drying rates $\dot{d}_{i}(t)$ at a distinct thickness $d_{i,\rm crit.}$ as defined by a critical concentration $c_{\rm crit.}$. In colloquial terms, one could say that the goal for the air knife is to follow the drying front, as it propagates along the length of the film from left to right. For simplicity, the drying dynamics were borrowed from previous works\cite{ternesDryingDynamicsSolutionProcessed2019,Ternes2023}. They describe the thickness decrease of a methylammonium lead iodide perovksite solution film with one volatile species, N,N-Dimethylformamide, and follow the equation:

\begin{equation}
d_{i}(t) = d_{p,i} + d_{s,i}W\left( \exp \left( - \frac{R_{\rm{solv}}(T)}{d_{s,i}} \int \limits _0 ^t \beta_i(t)dt + C_i \right)   \right)
\label{eq:dyn} 
\end{equation}

where $d_i(t)$ is the film thickness at position $i$, $W$ is the Lambert-W function, $d_{p,i}$ is the dry film thickness at position $i$,  $d_{s,i}=d_{p,i}\tilde{\rho}_p /\tilde{\rho}_{\rm{solv}}$ is the dry film thickness scaled with the molar densities of the solute $p$ and the solvent, $R_{\rm{Solv}}(T)$ is a solvent and temperature dependent constant and $C_i = \ln W^{-1}((d_{i}(t=0)-d_{p,i})/d_{s,i})$ is determined by the start condition.

\subsection{The air-knife and its mass transport coefficient}\label{sec:mass_transfer}

\begin{figure}[h!]
\centering

\includegraphics[width=0.6\textwidth]{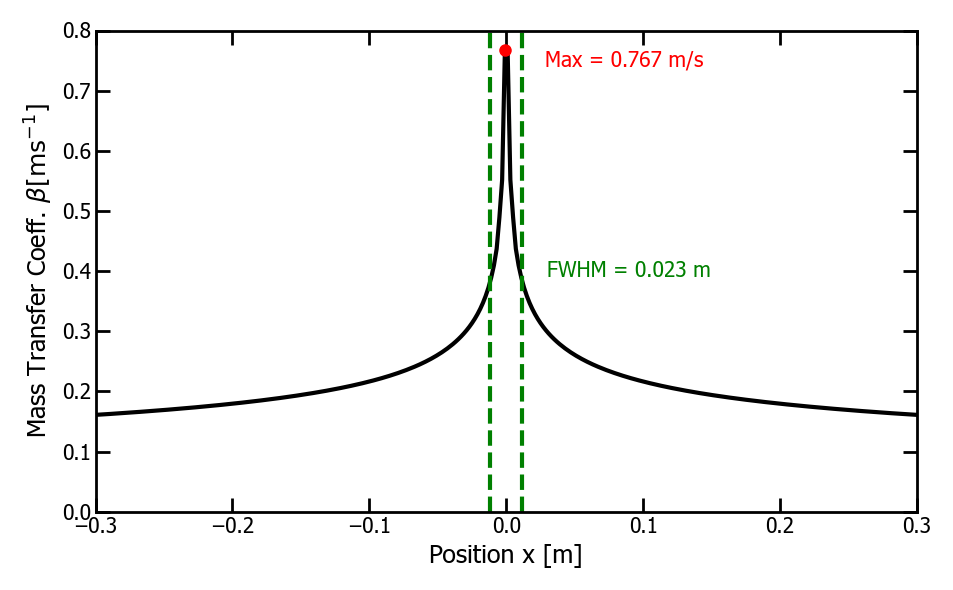}

\caption{Mass Transport coefficient $\beta (x)$ for the chosen DMF-based solution as calculated from the empirically validated Nusselt correlation by M. Nirmalkumar el al. \cite{nirmalkumarLocalHeatTransfer2011}. FWHM($\beta$) and $\beta_{\rm max}$ are indicated.}
\label{fig:beta}
\end{figure}

 Herein, the air knife is assumed to be a slot nozzle mounted at an angle of 90° above a flat plate. To estimate the spatial dependence of the mass transport coefficient $\beta$, a spatially resolved  correlation of Nusselt numbers by Nirmalkumar et al. is used \cite{nirmalkumarLocalHeatTransfer2011}. The correlation is defined  over a wide range of $x/b$, where $b$ is the width of the slot - starting from the stagnation region $0<x/b\leq2$, over the turbulent transition region $2<x/b\leq5$ toward the wall jet region $5<x/b$. Figure \ref{fig:beta} shows the mass transfer coefficient that was calculated for the correlation by leveraging the heat-mass transfer analogy and smoothing the transitions between the regions. As concrete exemplary parameters,  $h=1\,\rm{mm}$  (height over the sample), $b=1\,\rm{mm}$  (Slot Width), $u_0  =100\,\rm{m/s}$  (air flow speed) are chosen. However, it must be noted that the methodology presented herein can be applied to any spatially resolved function $\beta$.  The only assumption are that the mass transfer has a maximum at the nozzle center from which it falls off toward the edges.

\subsection{Goal of optimization of air-knife trajectory}

To optimize the drying process in the above described system, we need to define the goal of optimization. Due to the asymmetry of the setup explained above, it is not possible to achieve the same drying rate at every instant on every position of the film. However, there is a feature that can be exploited if the material system has a critical stage of drying at thickness $d_{i,\rm crit.}$, where drying rate impacts the result most. Note that this assumption was empirically shown to hold true for crystallization processes for perovskite solar cells\cite{majewskiSimulationPerovskiteThin2025,ternesDryingCoatingPerovskite2022,stargerFormationTrajectoriesSolutionprocessed2025}, who have a critical concentration at which crystallization starts, associated with a thickness $d_{i,\rm crit.}$. So, in the example herein, $d_{i,\rm crit.}$ is chosen very close to $d_{i,p}$ as it is the case for these perovksite thin films, but the parameter can be adjusted to any other drying stage.  The goal of optimization is the following: 

\begin{theorem*}
Maximize -$\langle \dot{d}_{i,\rm crit.} \rangle$ where the mean value is calculated over all positions.
\end{theorem*}
 
 The optimization goal is motivated by the idea that, in an ideal case scenario, the drying front should be synchronized with the position of the air knife center, where the mass transfer is at a maximum. At the same time, due to the existence of the maximu, achievable drying rate imposed by $\beta_{\rm{max}}$, the drying result will be consistent if the mean value is at a maximum. The above goal is adjustable in a sense that a different target drying rate as induced by $0<\beta(x_{\rm ref}) < \beta_{\rm max}$ could be chosen. However, if lower drying rates are needed, it is in general preferable to lower the air flow speed than to optimize drying at a value below $\beta_{max}$. The reason is that $\beta(x)$ at lower air speeds becomes more homogeneous and the film surface is less likely to be disturbed by the air pressure. Further note that in some cases the relative drying rate -$\dot{d}_i/d_i$ (which is identical to the  supersaturation rate\cite{ternesModelingFundamentalDynamics2024a}), might be more decisive than the absolute drying rate $\dot{d}_i$. In this case, the goal of optimization can be adjusted. However, for simplicity,  absolute drying rate is used herein because it showcases better the idea of keeping the air knife fully synchronized with the drying front at all times. In any case, for most scenarios, the thickness differences are small compared to the difference in $\dot{d}_i$ that is caused by movement of the air knife. Therefore, only minimal differences in the outcome are expected.

\subsection{Start scenarios of coated wet film thicknesses}

\begin{figure}[h!]
\centering

\includegraphics[width=1\textwidth]{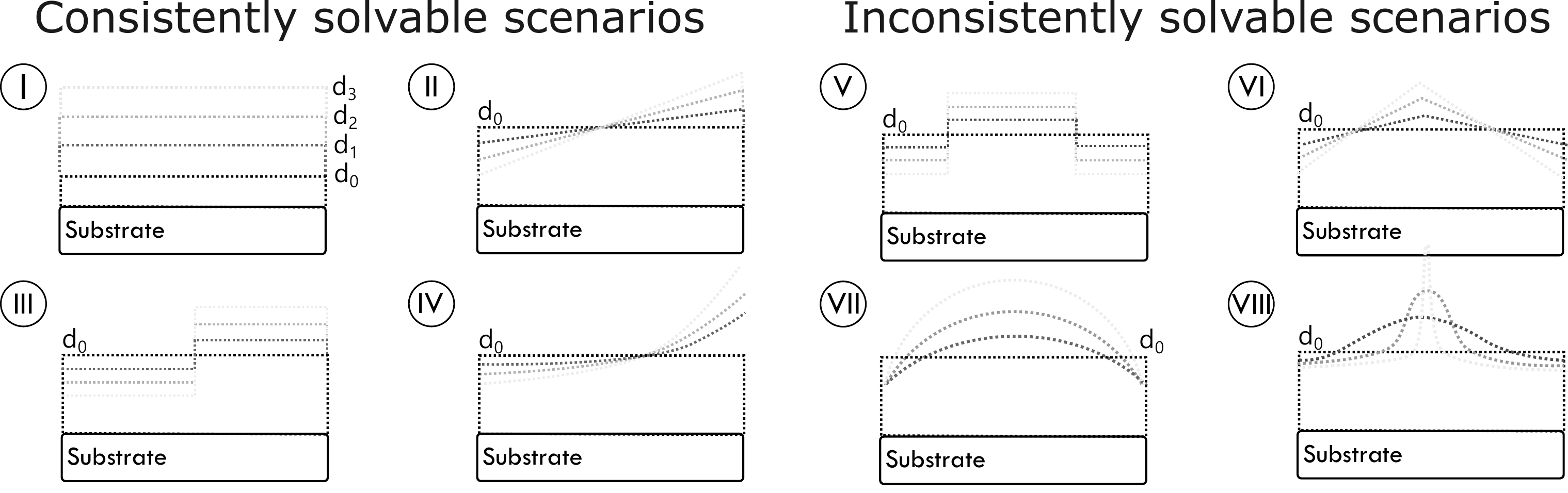}

\caption{Start scenarios for coated wet film thickness that will be dried in the simulated environment. All scenarios (I-IV) with a non-decreasing wet film thickness over the substrate width are consistently solvable, whereas the scenarios of a convex film thickness over the substrate (V-VIII) are potentially inconsistent. To make the scenarios comparable, the initial wet film thickness of the film is $d_0=4\,\rm{\mu m}$ is kept in every scenario and the volume of the thin films is kept constant throughout the variation.}
\label{fig:scenarios}
\end{figure}

Figure \ref{fig:scenarios} depicts the different scenarios of spatial distribution of deposited solution film thickness. None of the scenarios is physical in a sense that the film surface was calculated by realistic material parameters such as surface energies and mass densities. Instead, they are pure mathematical functions that showcase the impact of different film height variations on the optimization. More realistic distributions of film height over the sample can be studied in a future work.  Scenario I features a variation  of perfectly homogeneous solution films with different thicknesses. The other scenarios keep a perfect wet film with thickness $d_0 = 4\,\rm{\mu m}$ as a reference and then investigate the impact of different variations of film thickness over the sample area. The total volume of solution to be dried is however kept constant. Scenarios I-IV feature a non-decreasing film thickness over the substrate area. It will be shown below that this feature is decisive for a consistent set of equations. Scenarios V-VIII however feature a convex film thickness that increases initially and then decreases again. They are potentially inconsistent, meaning that the residuals do not always decrease down to machine precision, but the optimization result can still be useful. Later on, a modification of Scenarios V-VIII is presented where the same variation is applied in a way that the film surface has a concave form. These are denoted as V'-VIII'.

\subsection{Air knife-Trajectory optimization}\label{sec:greedy}

The principal idea for finding a solution to the problem posed above is that the points $x_i$ are approached by the air knife in succession. Thus, in the ideal case, the air knife drives from $x_{i-1}$ to $x_i$ with the velocity value $u_i$ adjusted such that the film thickness $d_i(t)$ hits $d_{i,\rm crit.}$ exactly when the nozzle reaches this position. Evidently, to calculate $u_i$, one needs to consider all $(u_1,...,u_{i-1})$  and the effect the previous dynamic drying of the air knife had on the thin film at position $i$. This can be done by transforming the integral in Equation \ref{eq:dyn} into the spatial domain and summing up all segments where $dt = -dx /u_k$ holds ($\beta(x)$ is integrated in negative $x$ direction from $x_k$ to $x_{k-1}$). This yields a set of $N$ equation of type

\begin{equation}
\frac{d_{i, \rm crit.}-d_{p,i}}{d_{s,i}} = W\left( \exp \left(C_i - \frac{R_{solv}}{d_{s,i}}  \sum_{k=1} ^i \frac{B(\Delta x (1+ i - k ))-B(\Delta x(i - k ))  }{u_{k}}  \right)\right)
\label{eq:main}
\end{equation}

where $B$ is the antiderivative of $\beta$ and $\Delta x=x_{i}-x_{i-1}$. Because $W(\exp (-x)) \approx \exp (-x)$ for large $x$ and  $d_{i, \rm crit.}$ was chosen very close to $d_{p,i}$, it is sensible to calculate the logarithm of Equation \ref{eq:main} to facilitate the gradient decent. We then calculate the residuals by subtracting the right side of the above equation from the left side as

\begin{equation}
r_{1,i}(d_{i, \rm crit.},d_{s,i},d_{0,i}) =  \ln\left( \texttt{LS} \left(d_{0,i},d_{s,i},d_{p,i}\right) \right) - \ln\left( \texttt{RS} \left( d_{0,i},d_{s,i}   \right)   \right)
\label{eq:resid}
\end{equation}

On the bases of these residuals, the Jacobian was calculated and a Levenberg–Marquardt gradient was performed with \emph{scipy.optimize.least\_squares} showing fast conversion to the optimal vector $\mathbf{u}_{\rm{opt,1}}$ (before performing the optimization, the admissible velocity values were confined within the interval $(0,1\,\rm{m/s})$ by using a sigmoid function and the Jacobian was adjusted accordingly). There is an issue regarding the residuals as described in Equation \ref{eq:resid}. The equation ensures that in the case $r_{1,i}(d_{i, \rm crit.},d_{s,i},d_{0,i})=0$, the air knife is situated exactly over the position $x_i$ when $d_i=d_{
i, \rm crit.}$. This is a satisfiable condition in consistent solutions. However, in some scenarios there is no consistent solution.  In these cases, $r_{1,i}$ does not scale correctly  with the error on the original optimization goal $\dot{d}_{i, \rm crit.}$. For example, it is clear that the true cost of not hitting $d_{i, \rm crit.}$ will depend on the current velocity of air knife.  To get the correct residual scaling for the  optimization target $\dot{d}_{i, \rm crit.}$, we perform a second optimization with 

\begin{equation}
r_{2,i}(d_{i, \rm crit.},d_{s,i},d_{0,i}) =  \frac{d_{i, \rm crit.}-d_{p,i}}{d_{i, \rm crit.}-d_{p,i}+d_{s,i}} - \frac{1}{1+1/\texttt{RS} \left( d_{0,i},d_{s,i}   \right)  }
\end{equation}

which is motivated from the known derivative of Equation $\ref{eq:dyn}$. Gradient decent does however not perform as well with these residuals. Therefore, $\mathbf{u}_{\rm{opt,1}}$ is passed as an initial guess to guide the gradient decent. By minimizing the squared residuals $r_{2,i}$, the mean drying rate averaged over all positions should be at a maximum, even in situations of inconsistent equations.

 \subsection{Numerical smoothing of air knife trajectory}
 
 For calculating the trajectory $\hat{x}(t)$ of the air knife, we calculate the times $t_i$ when the air knife reaches position $x_i$ as follows. 
 
 \begin{equation}
 t_i(x_i) = \int \limits _{0} ^{x_i} \frac{dx}{u(x)} = \sum_{k=1} ^i \frac{\Delta x}{u_k} 
 \label{eq:traj}
 \end{equation}
 
We thus obtain the vectors $(x_1,x_2...,x_N)$ and ($t_1,t_2,...,t_N$), where the spacing of the time points diminishes with $i$ due to the higher velocities in the end of the drying process. This bare simulation result assumes infinite accelerations to change velocity from $u_i$ to $u_{i+1}$. To smooth the trajectory and make it physically realistic, a windowed third degree polynomial fit is performed on the above data and the result is interpolated for equal spacing of data points. In this way, a regularized acceleration is ensured, while the variability in acceleration over time and accurate following of the required trajectory are still enabled. In a real-world scenario, a robot axis could be programmed with a feedback cycle to follow the required trajectory (to extend the trajectory a little further in time and space, we assume zero acceleration after the air knife has crossed the point $x_N$ and let $u_N$ be constant until $x>1.5 L_m$.).

\section{Results and Discussion}

\subsection{Trajectories of consistently solvable scenarios for achieving maximum absolute drying rate}

\begin{figure}[h!]
    \centering
    \begin{subfigure}{1\textwidth}
        \centering
        \includegraphics[width=\linewidth]{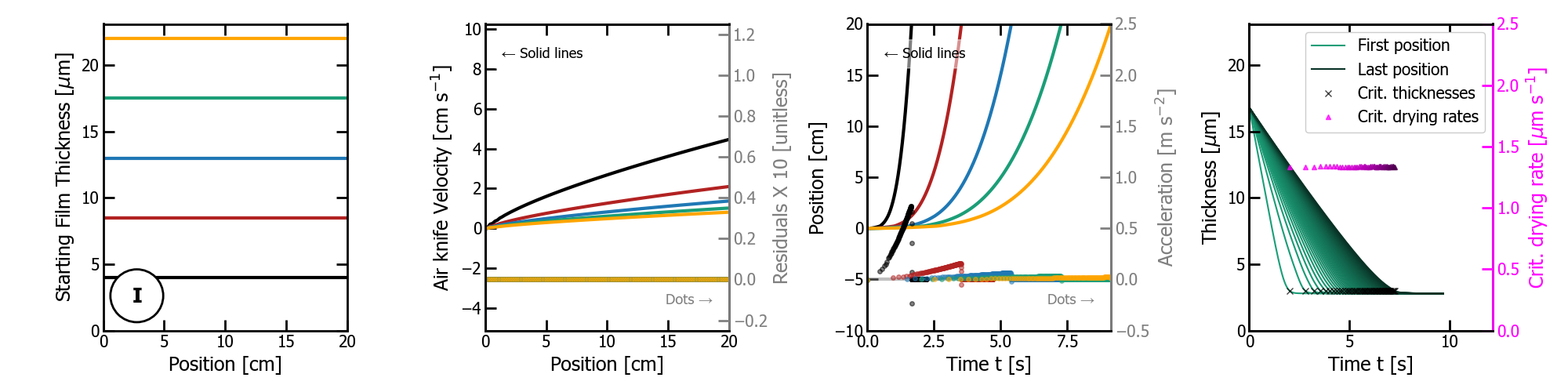}
    \end{subfigure}

    \begin{subfigure}{1\textwidth}
        \centering
        \includegraphics[width=\linewidth]{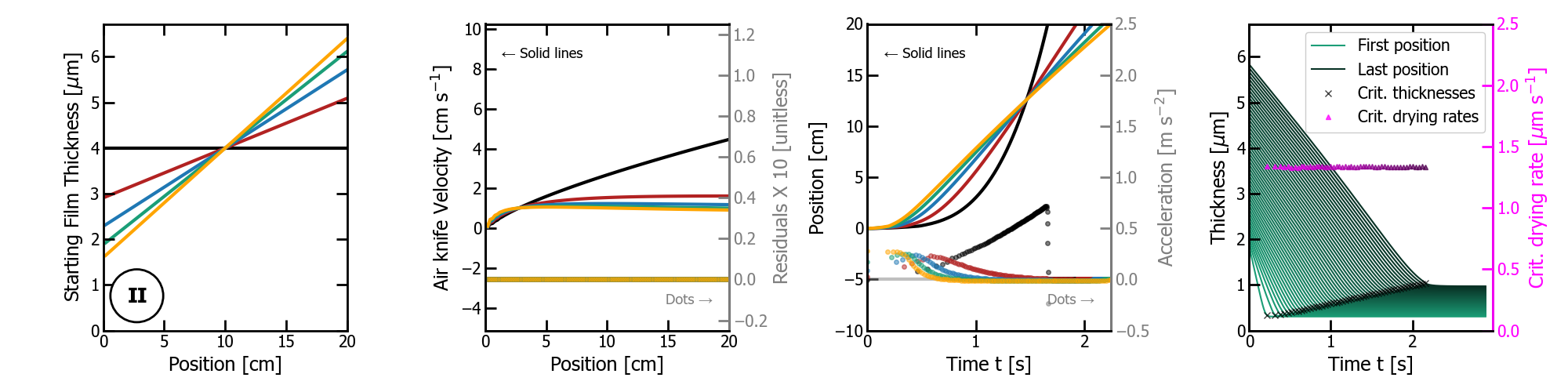}
    \end{subfigure}

    \begin{subfigure}{1\textwidth}
        \centering
        \includegraphics[width=\linewidth]{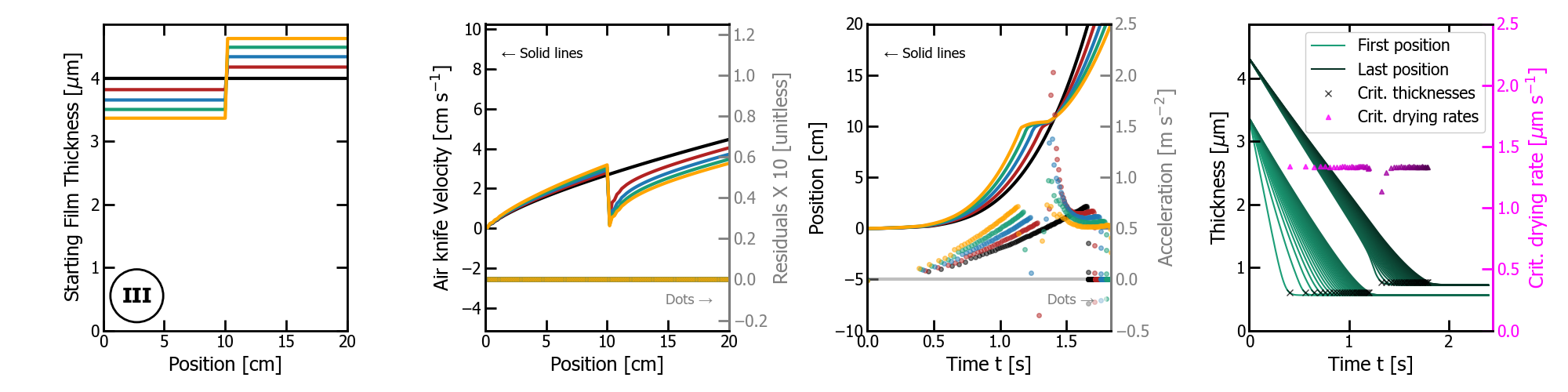}
    \end{subfigure}
    
    \begin{subfigure}{1\textwidth}
        \centering
        \includegraphics[width=\linewidth]{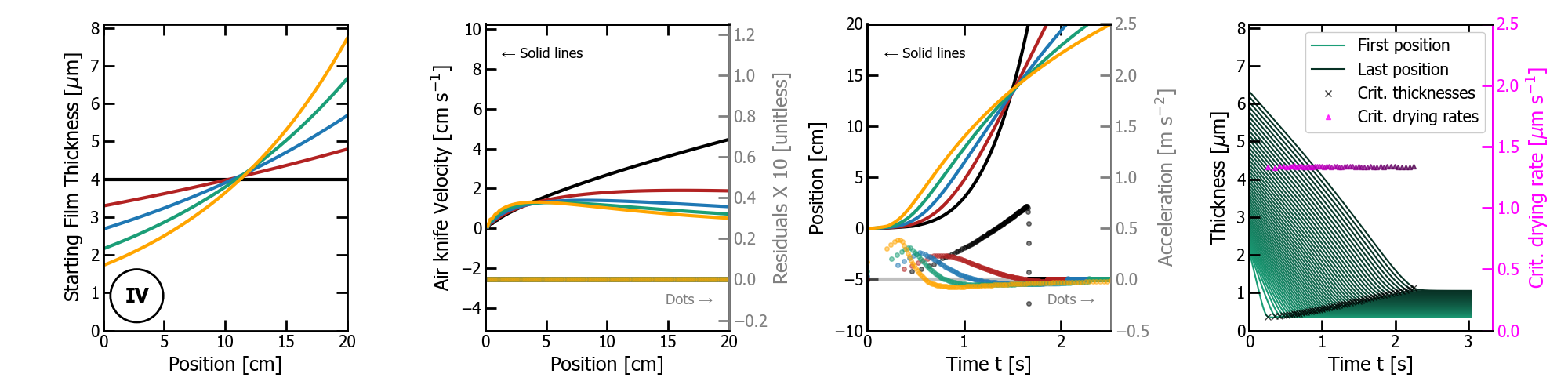}
    \end{subfigure}

    \caption{Ideal solutions of scenarios I-IV.  The first column of the left depicts the starting wet film thickness distribution, the second column the calculated air knife velocities (solid lines, left black axis) and the associated residuals (dots, right gray axis), the third column the smoothed air knife trajectory, $\hat{x}(t)$, (solid lines, left black axis) the associated acceleration $a(t)$ (dots, right gray axis) and the last column the thicknesses $d_i(t)$ over time corresponding to the turquoise line in the other plots along with the calculated drying rates (violet triangles) over the critical thicknesses (crosses). }
    \label{fig:ideal_scenarios}
\end{figure}

In Figure \ref{fig:ideal_scenarios}, the results of the optimization explained in Section \ref{sec:greedy} for the scenarios I-IV is depicted. In the first column from the left, we see the initial wet film thickness distributions for the scenarios  I-IV according to Figure \ref{fig:scenarios}. The second column depicts the resulting air knife velocities obtained by the gradient decent optimization plotted over the position on the substrate as solid lines (black axis on the left) and the associated residuals $r_{2,i}$ (gray axis on the right) as dots. For better comparability, the same y-axis scales were chosen for all plots in Figures \ref{fig:ideal_scenarios}-\ref{fig:non_ideal_scenarios_concave}. As a consequence, the randomly scattered residuals appear on the zero line in all of the scenarios I-IV (in detail they are smaller than $10^{-14}$), indicating a consistent solution. The third column shows the calculated trajectories over time,$\hat{x}(t)$, according to Equation \ref{eq:traj} as solid lines (left black axis) along with the calculated acceleration values as dots (right gray axis). Finally, the fourth column depicts the drying dynamics  that is the thickness decrease of the wet film over time for the turquoise solid line corresponding to the second highest variation of wet film thickness in every scenario. As a consistency check, the drying dynamics were calculated according Equation \ref{eq:dyn}, by using the trajectory $\hat{x}(t)$ from Equation \ref{eq:traj} to calculate the $\beta_i(t)=\beta(x_i-\hat{x}(t) )$ seen by each position $i$. The color of the lines darkens with increasing position on the substrate. On the right axis (magenta) the critical drying rate (calculated directly from the drying dynamics) is plotted as triangles of magenta color that darken as well with the position. Also the critical thicknesses, $d_{i,\rm crit.}$, are indicated on every drying curve as crosses. It is clearly visible that, indeed, all the drying rates are constant over the substrate length within the tolerance of the numerical error (one exception is detailed below). Further, the moment in time, when the last drying curve hits $d_{N,\rm crit.}$ coincides with the moment when the air knife trajectory reaches the substrate edge. These considerations show the consistency of the results.

Focusing on scenario I, we find that, consistent with intuition, the ideal air knife velocities increase with decreasing film thickness because thinner films dry faster. For all thickness scenarios, the ideal air knife speed over the position increases monotonously. When analyzing the acceleration values required for achieving the trajectory of the air nozzle, we find the need to linearly increase acceleration over time (constant jerk). This is an achievable constraint (at the predicted order of magnitudes) for state-of-the-art robot axes. Note that, to the author's knowledge, these conditions are not met in any experimental setup used for drying of solution thin films on similarly sized substrates. Instead, constant air knife speeds or a static air knife are typically employed\cite{wuProgressBladecoatingMethod2021,subbiahEnhancingPerformanceBladeCoated2024,geistertControllingThinFilm2023a,ternesDryingCoatingPerovskite2022,chenBladeCoatedPerovskitesTextured2020}. Consequently, there is a large potential for using accelerated air knives with optimal trajectories adapted to the deposited wet film thickness opening an experimental space for exploration. The drying times over the substrate size are not constant, which becomes clear by the fact that drying starts at one substrate edge, while already having an effect on the film at the other edge. So, drying accelerates as the air-knife is moved consecutively over the substrate. Toward the last position on the substrate, the spacing of the drying dynamics reduces due to the faster movement of the air knife, while keeping the drying rates at critical thickness at a constant level as required.

Scenario II assumes a linearly increasing slope of the deposited wet film thickness. Interestingly, such a slope balances the required air knife velocity increase seen previously in scenario I. Depending on the magnitude of the slope, the required velocity profile either increases more slowly or undergoes a shallow maximum. Accordingly, the required acceleration values fall off toward zero after the initial acceleration. This feature can be beneficial when drying blade-coated films that typically have increasing film thickness along the coating direction. If these irregular films are dried at the required air knife velocity of the shallow maximum, large portions of the film are dried consistently (that is at constant absolute drying rate). This could give an explanation why air-knife drying works well on blade-coated thin films in a lot of scenarios for perovskite solar cell manufacturing. The drying dynamics appear equally spaced at higher thicknesses and do not cross. This implies that the drying front moves approximately at constant speed, while remaining synchronized with the air knife. In other words, the longer drying times of the thicker films are counterbalanced by the fact that the nozzle needs more time to reach these positions.

Scenario III assumes a discrete jump in the film thickness to be dried (this is clearly not physical, but interesting as an academic example). As a consequence,  the required velocity profile experiences a discrete jump, as well. The air knife velocity at the jump must be re-adjusted to match the increases film thickness due to the step. This scenario could be an interesting case study for intermittent coating or coating on uneven surfaces where sudden thickness changes can occur. The smoothing of the air knife trajectory then causes two points at the junction to be slightly sub-optimally dried.

Scenario IV assumes an exponentially increasing film thickness as opposed to the linear case shown in scenario II. A similar velocity maximum appears as with the linearly increasing film thickness, but the maximum in velocity is now more pronounced. Required acceleration shifts to the negative slowly after the initial acceleration and the drying dynamics get shifted more and more with increasing position. This scenario illustrates that knowing the exact shape of the film to be dried can have an impact on the homogeneity of the drying process on a wafer-sized substrate.

\pagebreak

\subsection{Trajectories of non-ideal scenarios}

\begin{figure}[h!]
    \centering
    \begin{subfigure}{1\textwidth}
        \centering
        \includegraphics[width=\linewidth]{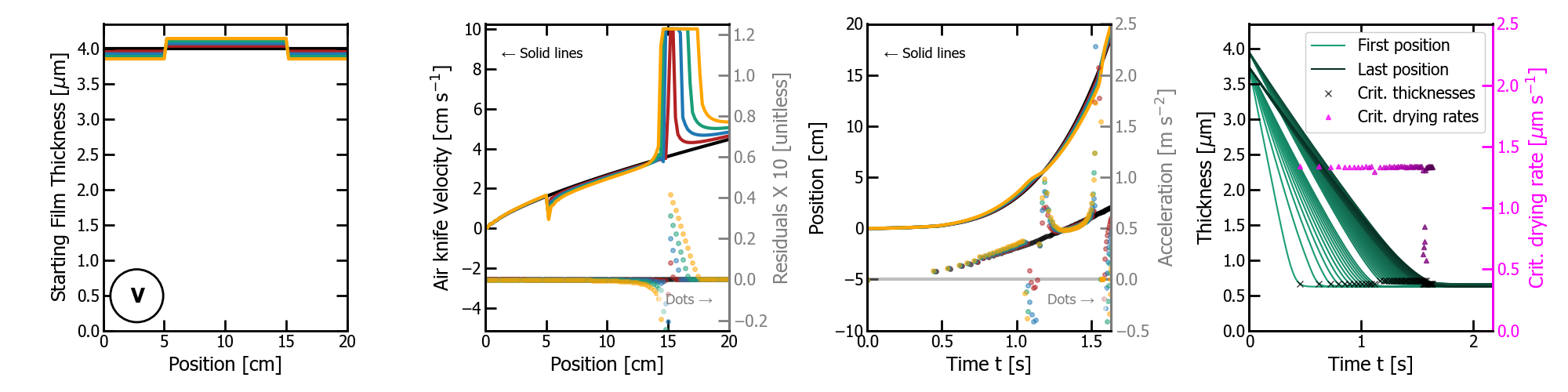}
    \end{subfigure}

    \begin{subfigure}{1\textwidth}
        \centering
        \includegraphics[width=\linewidth]{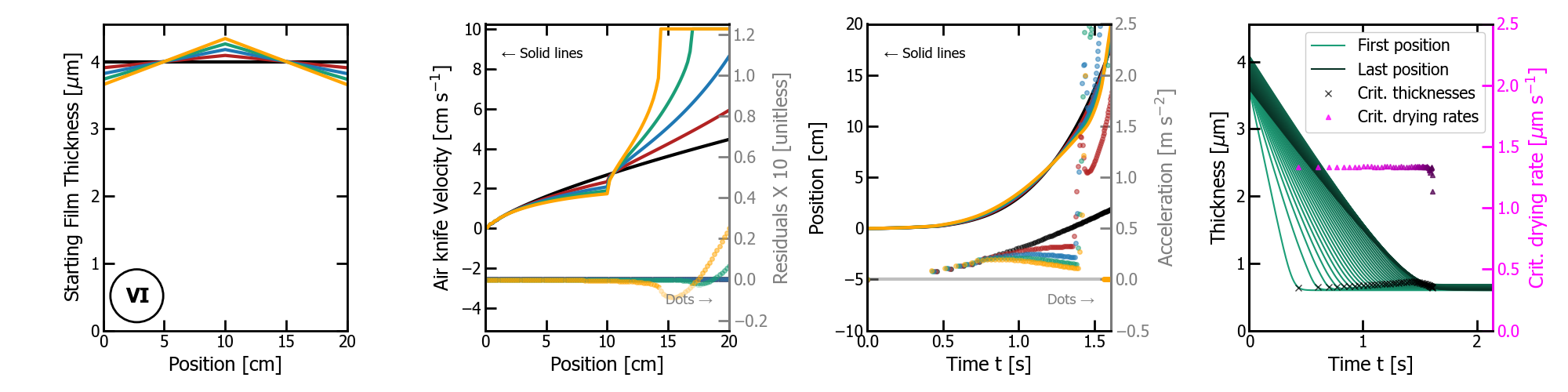}
    \end{subfigure}

    \begin{subfigure}{1\textwidth}
        \centering
        \includegraphics[width=\linewidth]{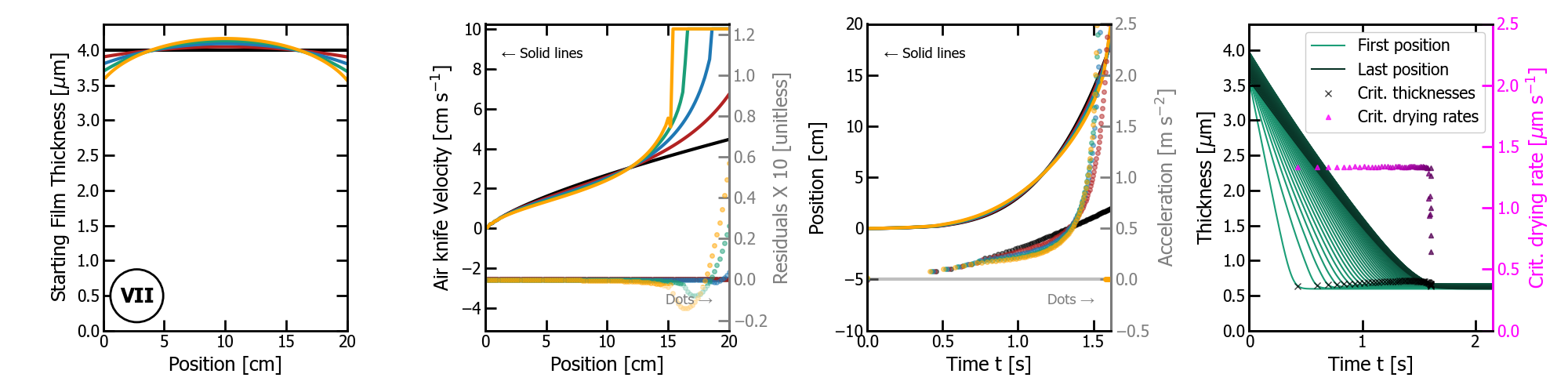}
    \end{subfigure}
    
    \begin{subfigure}{1\textwidth}
        \centering
        \includegraphics[width=\linewidth]{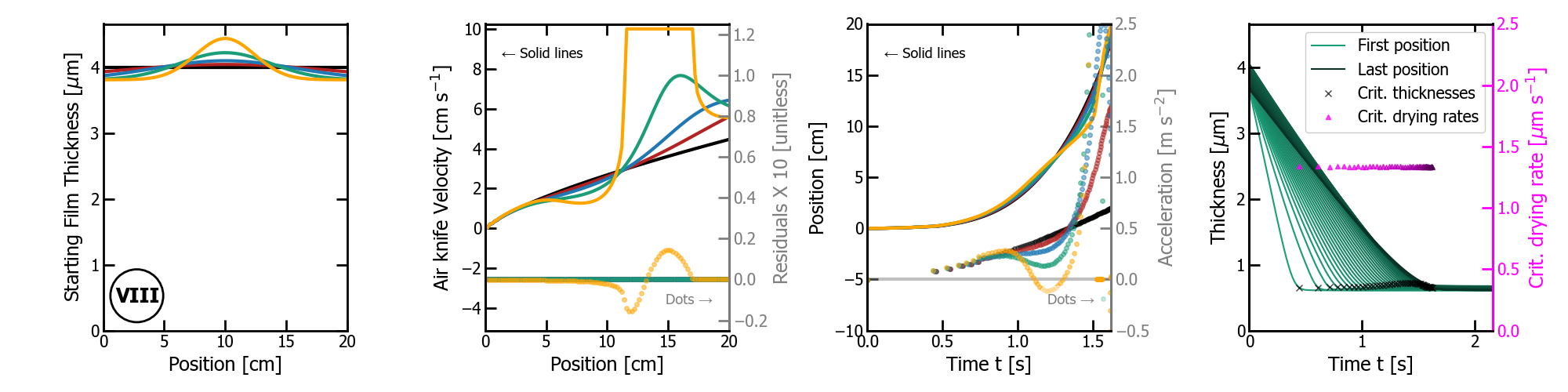}
    \end{subfigure}

    \caption{Optimal solutions of scenarios V-VIII.  The first column of the left depicts the starting wet film thickness distribution, the second column the calculated air knife velocities (solid lines, left black axis) and the associated residuals (dots, right gray axis), the third column the smoothed air knife trajectory, $\hat{x}(t)$, (solid lines, left black axis) the associated acceleration $a(t)$ (dots, right gray axis) and the last column the thicknesses $d_i(t)$ over time corresponding to the turquoise line in the other plots along with the calculated drying rates (violet triangles) over the critical thicknesses (crosses). }
    \label{fig:non_ideal_scenarios}
\end{figure}

The results for the simulation of the scenarios V-VIII for convex film shapes are depicted in Figure \ref{fig:non_ideal_scenarios}. All of the scenarios share a common property, which is that the wet film thicknesses follow a convex shape over the substrate area. Conclusively, the results of the optimization exhibit a similar pattern, as well, with some particularities in the end region of the substrate. For drying these convex  wet film consistently, just as in scenarios I-IV, the air knife will increase in velocity, at first, and then slightly reduce the velocity to account for the thicker film in the substrate center. As visible in the residuals (see second column dots and gray axis), on the second half of the substrate where the wet film thickness starts decreasing, the drying rates deviate from the optimum. The first deviation goes  downward, indicating that the speed of the air knife is slightly higher than required to allow for a more consistent drying of the last part. Then, the residuals shift upward, indicating that the air knife is not fast enough to follow  the drying front accurately. This is where the previously imposed boundary  of $1\,\rm{m/s}$ of air knife speed comes into play to regularize the inconsistent solution and keep the parameters in a physically feasible regime.

In general, consistent with intuition, the last part of the substrate where the wet film thickness decreases again, is the most challenging to handle. This has two reasons: 1) these parts will have a higher likelihood of already exceeding $d_{i, \rm crit.}$ before the nozzle reaches them and 2) the nozzle already moves much quicker when it reaches these points, so deviations from the optimal speed cause more loss in achieved drying rate.  In those scenarios where the lateral slope of the wet film thickness decrease becomes smaller at the end of the substrate (V and VIII), the nozzle will catch up again with the drying front and dry the film consistently in the end region. 

\begin{figure}[h!]
    \centering
    \begin{subfigure}{1\textwidth}
        \centering
        \includegraphics[width=\linewidth]{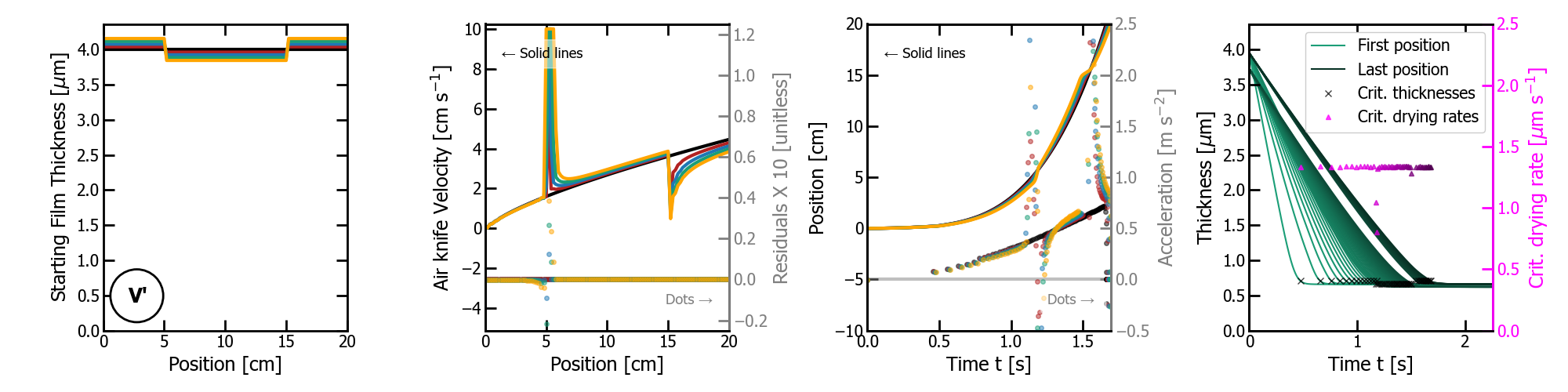}
    \end{subfigure}

    \begin{subfigure}{1\textwidth}
        \centering
        \includegraphics[width=\linewidth]{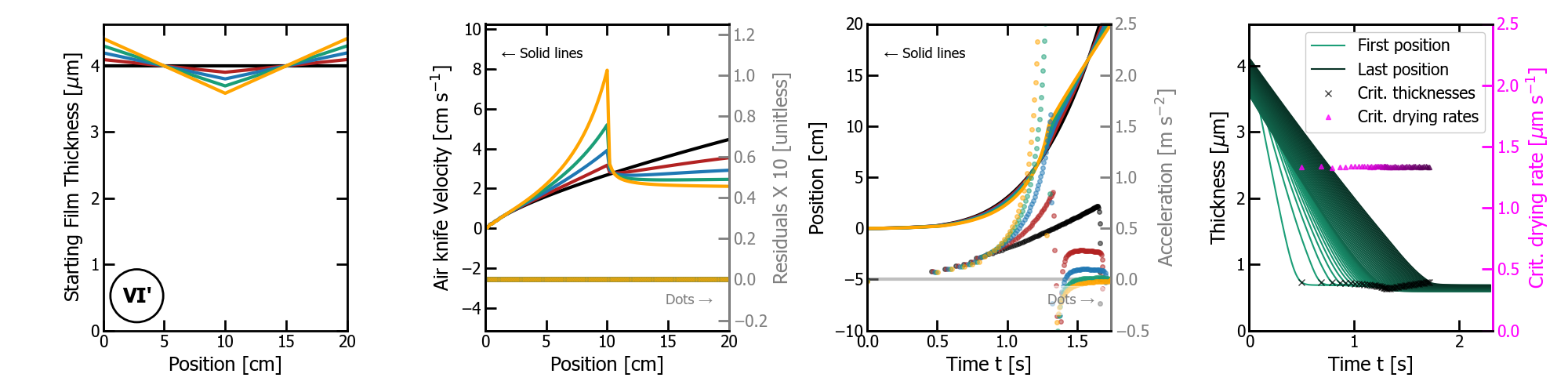}
    \end{subfigure}

    \begin{subfigure}{1\textwidth}
        \centering
        \includegraphics[width=\linewidth]{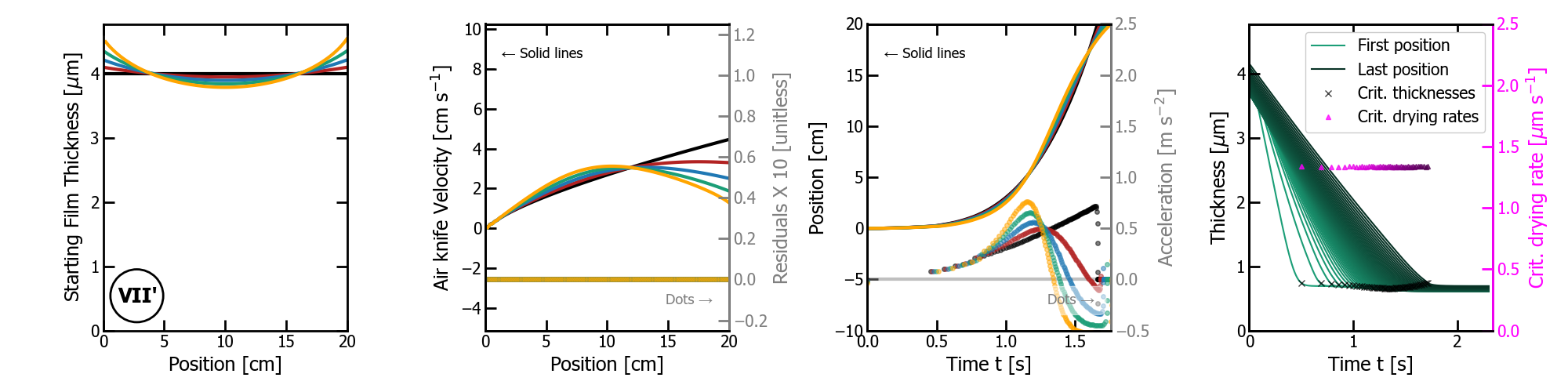}
    \end{subfigure}
    
    \begin{subfigure}{1\textwidth}
        \centering
        \includegraphics[width=\linewidth]{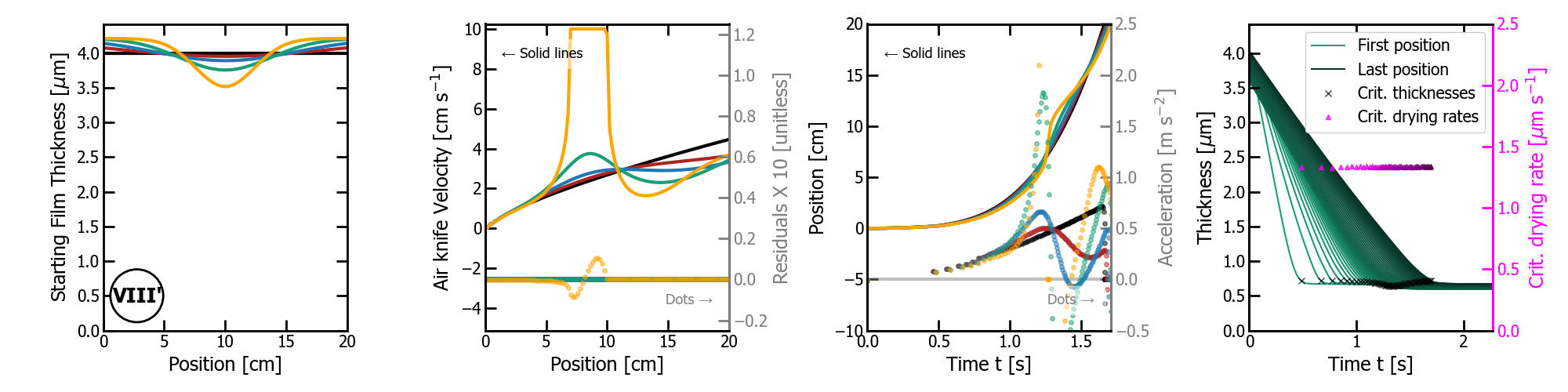}
    \end{subfigure}
    \caption{Optimal solutions of scenarios V'-VIII' for concave film shapes.  The first column of the left depicts the starting wet film thickness distribution, the second column the calculated air knife velocities (solid lines, left black axis) and the associated residuals (dots, right gray axis), the third column the smoothed air knife trajectory, $\hat{x}(t)$, (solid lines, left black axis) the associated acceleration $a(t)$ (dots, right gray axis) and the last column the thicknesses $d_i(t)$ over time corresponding to the turquoise line in the other plots along with the calculated drying rates (violet triangles) over the critical thicknesses (crosses). }
    \label{fig:non_ideal_scenarios_concave}
\end{figure}

The results for the simulation of the scenarios V'-VIII' for concave film shapes are depicted in Figure \ref{fig:non_ideal_scenarios_concave}. As compared to the convex film shapes, the drying of concave films is easier to control. This is due to the fact that the lowest film thickness occurs already in the middle of the substrate, which is reached earlier in the process. The subsequently increasing film thickness at the end part of the  substrate can then be dried consistently by simply reducing the velocity of the air knife again. 

\pagebreak

\pagebreak

\section{Conclusion and outlook}

This work details on how to achieve more consistent drying results when using a linearly moving air knife to dry a solution film on a substrate of medium size ($20\, \rm cm$ are chosen here, but the results apply to any situation where the substrate is wide several tens of FWHMs of the mass transfer coefficient distribution). As opposed to driving the substrate under the air knife with a constant speed, researchers and technologists should consider driving with non-zero acceleration over time. If jumps large jumps in wet film thickness occur, the air-knife velocity should be re-adjusted at these positions. To calculate an optimal trajectory, it is crucial to have precise knowledge on 1) the lateral wet film thickness over the substrate area and 2) the lateral inhomogneiety of the mass transfer coefficient. If the film thickness has a gradient, it should always be dried starting from the thinner side. Gradients help, in fact, the stabilization of drying rates and reduce the required acceleration values (or rates). If the film thickness has a convex or concave form over the substrate area, there is still a merit in optimizing the air knife trajectory although the result may not be fully consistent. Depending on the magnitude of the thickness variation with respect to the shape of the exerted mass transfer, this may, however, require very high acceleration values at the end of the drying process. A possible mitigation strategy would be to dry the film with two air knives approaching from both substrate edges toward the inner part of the substrate.

In future works, this knowledge should be applied to a real-world drying setup under highly controlled conditions to test the predicted dynamics. In this case, the precision, tolerances and repeatably of both coating and drying processes would need to be considered to apply the results showcased herein. Further, more parameters such as the gas speed could be optimized in parallel to the air knife trajectory to obtain more consistent results on very irregular thin film thicknesses. The optimization framework could further be extended to include a correct description of surface tensions and the deformation of the film surface.

\section{Acknoweldgements}

S.T. acknowledges the European Union's Framework Programme for Research and Innovation Horizon Europe (2021-2027) under the Marie Skłodowska-Curie Grant Agreement No. 101107885 “INT-PVK-PRINT”.

\bibliographystyle{unsrt}
\bibliography{references}

\begin{thebibliography}{10}

\bibitem{Heyd25}
Rodolphe Heyd, Julie Fichot, Driss Lahboub, Abderrahim Bakak, Christophe
  Josserand, and Marie-Louise Saboungi.
\newblock Drying of aqueous films, an application of heat and mass transfer.
\newblock In Diana Enescu, editor, {\em Heat and Mass Transfer - from
  Fundamentals to Advanced Applications}, chapter~8. IntechOpen, London, 2025.

\bibitem{buttThinFilmCoatingMethods2022}
Muhammad~A. Butt.
\newblock Thin-{{Film Coating Methods}}: {{A Successful Marriage}} of
  {{High-Quality}} and {{Cost-Effectiveness}}---{{A Brief Exploration}}.
\newblock {\em Coatings}, 12(8):1115, August 2022.

\bibitem{niskanenMonitoringDryingProcess2019}
Ilpo Niskanen, Janne Lauri, Jukka R{\"a}ty, Rauno Heikkil{\"a}, Henrikki
  Liimatainen, Taro Hashimoto, Tapio Fabritius, Kaitao Zhang, and Masayuki
  Yokota.
\newblock Monitoring drying process of varnish by immersion solid matching
  method.
\newblock {\em Progress in Organic Coatings}, 136:105299, November 2019.

\bibitem{clementRecentAdvancesPrinted2022a}
Benoit Clement, Miaoqiang Lyu, Eeshan Sandeep~Kulkarni, Tongen Lin, Yuxiang Hu,
  Vera Lockett, Chris Greig, and Lianzhou Wang.
\newblock Recent {{Advances}} in {{Printed Thin-Film Batteries}}.
\newblock {\em Engineering}, 13:238--261, June 2022.

\bibitem{choolaeiRecentAdvancesChallenges2023}
Mohammadmehdi Choolaei, Mohsen~Fallah Vostakola, and Bahman~Amini Horri.
\newblock Recent {{Advances}} and {{Challenges}} in {{Thin-Film Fabrication
  Techniques}} for {{Low-Temperature Solid Oxide Fuel Cells}}.
\newblock {\em Crystals}, 13(7):1008, July 2023.

\bibitem{verbovenPrintingFlexibleLight2021}
I.~Verboven and W.~Deferme.
\newblock Printing of flexible light emitting devices: {{A}} review on
  different technologies and devices, printing technologies and
  state-of-the-art applications and future prospects.
\newblock {\em Progress in Materials Science}, 118:100760, May 2021.

\bibitem{toshimaRecentProgressOrganic2017}
Naoki Toshima.
\newblock Recent progress of organic and hybrid thermoelectric materials.
\newblock {\em Synthetic Metals}, 225:3--21, March 2017.

\bibitem{zhangRecentProgressEmerging2022}
Qichun Zhang, Wenping Hu, Henning Sirringhaus, and Klaus M{\"u}llen.
\newblock Recent {{Progress}} in {{Emerging Organic Semiconductors}}.
\newblock {\em Advanced Materials}, 34(22):2108701, 2022.

\bibitem{howardCoatedPrintedPerovskites2019}
Ian~A. Howard, Tobias Abzieher, Ihteaz~M. Hossain, Helge Eggers, Fabian
  Schackmar, Simon Ternes, Bryce~S. Richards, Uli Lemmer, and Ulrich~W.
  Paetzold.
\newblock Coated and {{Printed Perovskites}} for {{Photovoltaic Applications}}.
\newblock {\em Advanced Materials}, 31(26):1806702, 2019.

\bibitem{fanEfficientPreciseSolutionvacuum2025}
Yingping Fan, Zhixiao Qin, Lei Lu, Ni~Zhang, Yugang Liang, Shaowei Wang, Wenji
  Zhan, Jiahao Guo, Haifei Wang, Yuetian Chen, Yanfeng Miao, and Yixin Zhao.
\newblock An efficient and precise solution-vacuum hybrid batch fabrication of
  {{2D}}/{{3D}} perovskite submodules.
\newblock {\em Nature Communications}, 16(1):7019, July 2025.

\bibitem{sondergaardRolltoRollFabricationLarge2013}
Roar~R. S{\o}ndergaard, Markus H{\"o}sel, and Frederik~C. Krebs.
\newblock Roll-to-{{Roll}} fabrication of large area functional organic
  materials.
\newblock {\em Journal of Polymer Science Part B: Polymer Physics},
  51(1):16--34, 2013.

\bibitem{parvazianRolltorollRevolutionTackle2024}
Ershad Parvazian and Trystan Watson.
\newblock The roll-to-roll revolution to tackle the industrial leap for
  perovskite solar cells.
\newblock {\em Nature Communications}, 15(1):3983, May 2024.

\bibitem{dingReviewOperatingLimits2016}
Xiaoyu Ding, Jianhua Liu, and Tequila A.~L. Harris.
\newblock A review of the operating limits in slot die coating processes.
\newblock {\em AIChE Journal}, 62(7):2508--2524, 2016.

\bibitem{parkPracticalOperationsIntermittent2022}
Jin~Seok Park, Sanghun Jee, Byoungjin Chun, and Hyun~Wook Jung.
\newblock Practical operations for intermittent dual-layer slot coating
  processes.
\newblock {\em Korea-Australia Rheology Journal}, 34(3):181--186, August 2022.

\bibitem{diehmEdgeFormationHighSpeed2020}
Ralf Diehm, Hannes Weinmann, Jana Kumberg, Marcel Schmitt, J{\"u}rgen
  Fleischer, Philip Scharfer, and Wilhelm Schabel.
\newblock Edge {{Formation}} in {{High-Speed Intermittent Slot-Die Coating}} of
  {{Disruptively Stacked Thick Battery Electrodes}}.
\newblock {\em Energy Technology}, 8(2):1900137, 2020.

\bibitem{schackmarPerovskiteSolarCells2021}
Fabian Schackmar, Helge Eggers, Markus Frericks, Bryce~S. Richards, Uli Lemmer,
  Gerardo {Hernandez-Sosa}, and Ulrich~W. Paetzold.
\newblock Perovskite {{Solar Cells}} with {{All-Inkjet-Printed Absorber}} and
  {{Charge Transport Layers}}.
\newblock {\em Advanced Materials Technologies}, 6(2):2000271, 2021.

\bibitem{geistertControllingThinFilm2023a}
Kristina Geistert, Simon Ternes, David~B. Ritzer, and Ulrich~W. Paetzold.
\newblock Controlling {{Thin Film Morphology Formation}} during {{Gas
  Quenching}} of {{Slot-Die Coated Perovskite Solar Modules}}.
\newblock {\em ACS Applied Materials \& Interfaces}, 15(45):52519--52529,
  November 2023.

\bibitem{ternesDryingCoatingPerovskite2022}
Simon Ternes, Jonas Mohacsi, Nico L{\"u}dtke, H.~Minh Pham, Meri{\c c} Arslan,
  Philip Scharfer, Wilhelm Schabel, Bryce~S. Richards, and Ulrich~W. Paetzold.
\newblock Drying and {{Coating}} of {{Perovskite Thin Films}}: {{How}} to
  {{Control}} the {{Thin Film Morphology}} in {{Scalable Dynamic Coating
  Systems}}.
\newblock {\em ACS Applied Materials \& Interfaces}, 14(9):11300--11312, March
  2022.

\bibitem{ternesDryingDynamicsSolutionProcessed2019}
Simon Ternes, Tobias B{\"o}rnhorst, Jonas~A. Schwenzer, Ihteaz~M. Hossain,
  Tobias Abzieher, Waldemar Mehlmann, Uli Lemmer, Philip Scharfer, Wilhelm
  Schabel, Bryce~S. Richards, and Ulrich~W. Paetzold.
\newblock Drying {{Dynamics}} of {{Solution-Processed Perovskite Thin-Film
  Photovoltaics}}: {{In Situ Characterization}}, {{Modeling}}, and {{Process
  Control}}.
\newblock {\em Advanced Energy Materials}, 9(39):1901581, 2019.

\bibitem{majewskiSimulationPerovskiteThin2025}
M.~Majewski, S.~Qiu, O.~Ronsin, L.~L{\"u}er, V.~M.~Le Corre, T.~Du,
  C.~J.~Brabec, H.-J. Egelhaaf, and J.~Harting.
\newblock Simulation of perovskite thin layer crystallization with varying
  evaporation rates.
\newblock {\em Materials Horizons}, 12(2):555--564, 2025.

\bibitem{stargerFormationTrajectoriesSolutionprocessed2025}
Jesse~L. Starger, Amy~E. Louks, Kelly Schutt, E.~Ashley Gaulding, Robert~W.
  Epps, Rosemary~C. Bramante, Ross~A. Kerner, Kai Zhu, Joseph~J. Berry,
  Nicolas~J. Alvarez, Richard~A. Cairncross, and Axel~F. Palmstrom.
\newblock Formation trajectories of solution-processed perovskite thin films
  from mixed solvents.
\newblock {\em Cell Reports Physical Science}, 6(7):102655, July 2025.

\bibitem{ternesModelingFundamentalDynamics2024a}
Simon Ternes, Felix Laufer, and Ulrich~W. Paetzold.
\newblock Modeling and {{Fundamental Dynamics}} of {{Vacuum}}, {{Gas}}, and
  {{Antisolvent Quenching}} for {{Scalable Perovskite Processes}}.
\newblock {\em Advanced Science}, 11(14):2308901, 2024.

\bibitem{guerrierDryingKineticsPolymer1998}
B{\'e}atrice Guerrier, Charles Bouchard, Catherine Allain, and Christine
  B{\'e}nard.
\newblock Drying kinetics of polymer films.
\newblock {\em AIChE Journal}, 44(4):791--798, 1998.

\bibitem{Ternes2023}
Simon Ternes.
\newblock In situ characterization and modelling of drying dynamics for
  scalable printing of hybrid perovskite photovoltaics.
\newblock {\em KIT Scientific Publishing}, March 2023.

\bibitem{nirmalkumarLocalHeatTransfer2011}
M.~Nirmalkumar, Vadiraj Katti, and S.~V. Prabhu.
\newblock Local heat transfer distribution on a smooth flat plate impinged by a
  slot jet.
\newblock {\em International Journal of Heat and Mass Transfer},
  54(1):727--738, January 2011.

\bibitem{wuProgressBladecoatingMethod2021}
Runsheng Wu, Chunhua Wang, Minhua Jiang, Chengyu Liu, Dongyang Liu, Shuigen Li,
  Qingrong Kong, Wei He, Changjun Zhan, Fayun Zhang, Xiaohong Liu, Bingchu
  Yang, and Wei Hu.
\newblock Progress in blade-coating method for perovskite solar cells toward
  commercialization.
\newblock {\em Journal of Renewable and Sustainable Energy}, 13(1):012701,
  February 2021.

\bibitem{subbiahEnhancingPerformanceBladeCoated2024}
Anand~S. Subbiah, Luis~V. Torres~Merino, Anil~R. Pininti, Vladyslav Hnapovskyi,
  Subhashri Mannar, Erkan Aydin, Arsalan Razzaq, Thomas~G. Allen, and Stefaan
  De~Wolf.
\newblock Enhancing the {{Performance}} of {{Blade-Coated
  Perovskite}}/{{Silicon Tandems}} via {{Molecular Doping}} and {{Interfacial
  Energy Alignment}}.
\newblock {\em ACS Energy Letters}, 9(2):727--731, February 2024.

\bibitem{chenBladeCoatedPerovskitesTextured2020}
Bo~Chen, Zhengshan~J. Yu, Salman Manzoor, Shen Wang, William Weigand, Zhenhua
  Yu, Guang Yang, Zhenyi Ni, Xuezeng Dai, Zachary~C. Holman, and Jinsong Huang.
\newblock Blade-{{Coated Perovskites}} on {{Textured Silicon}} for
  26\%-{{Efficient Monolithic Perovskite}}/{{Silicon Tandem Solar Cells}}.
\newblock {\em Joule}, 4(4):850--864, April 2020.

\end{thebibliography}

\end{document}